\documentstyle{article}

\newcommand{\be}{\begin{equation}}
\newcommand{\ee}{\end{equation}}
\newcommand{\bea}{\begin{eqnarray}}
\newcommand{\eea}{\end{eqnarray}}
\newcommand{\lb}{\label}
\begin{document}
\begin{titlepage}
\begin{flushright}
ZU-TH 25/93\\
\end{flushright}
\begin{center}
\vfill
{\large\bf  Quantum Gravity and Non-unitarity in Black Hole
Evaporation}
\vfill
{\bf Claus Kiefer}\footnote{Address after October, 1st: Institute for
Theoretical Physics, University of Freiburg, Hermann-Herder-Str. 3,
D-79104 Freiburg, Germany.}

\vskip 0.4cm
Institute for Theoretical Physics, University of Z\"urich,
Sch\"onberggasse 9,\\CH-8001 Z\"urich, Switzerland
\vskip 0.7cm
{\bf Rainer M\"uller}
\vskip 0.4cm
Fakult\"at f\"ur Physik, Universit\"at Konstanz, Postfach 5560 M
674,\\
D-78434 Konstanz, Germany
\vskip 0.7cm
{\bf Tejinder P. Singh}
\vskip 0.4cm
Tata Institute of Fundamental Research, Homi Bhabha Road,\\
Bombay 400005, India

\end{center}

\vfill

\begin{center}
{\bf Abstract}
\end{center}
\begin{quote}
We discuss the relevance of quantum gravitational corrections to the
functional
Schr\"odinger equation for the information loss paradox in black hole
evaporation. These corrections are found from the Wheeler-DeWitt
equation
through a semiclassical expansion scheme. The dominant contribution
in the final evaporation stage, when the black hole approaches the
Planck
regime, is a term which explicitly violates unitarity in the
non-gravitational sector. While pure states remain pure, there is
an increase in the degree of purity for non-pure states
in this sector. This result  holds
irrespective of whether full quantum gravity respects unitarity or
not.
\end{quote}

\vfill

\end{titlepage}

Soon after the discovery of black hole radiance, it was realized
that the semiclassical description of an evaporating black hole
seems to be at variance with the quantum mechanical unitarity of
time evolution \cite{Hawking76}. Let us consider a black hole that is
formed by the collapse of a matter distribution which we assume to be
in a
pure state initially. It will emit thermal radiation, leading to its
subsequent evaporation. If the evaporation is complete, only
the Hawking radiation is left behind. Since thermal radiation
cannot be described by a pure state, there seems to be an
evolution from pure into mixed states, which is connected with a loss
of
information about the system. This constitutes the black hole
information loss paradox which has received considerable attention
recently
(see \cite{Page93,Preskill92} for a review).

A number of possibilities have been proposed to resolve the
paradox :
(1) The above description is incomplete and the true time
evolution is unitary \cite{Page80}. This would mean that
information must be encoded in the black hole radiation,
for example by stimulated emission
\cite{Mueller93,Bekenstein93,Schiffer93} or in correlations
\cite{Carlitz87,Wilczek93}. It seems unlikely, however, that
any of the mechanisms proposed up to now will be able to restore
the unitarity of the black hole evaporation.
(2) The black hole evaporation is not complete but ceases at a mass
of
 the order of the
Planck mass, leaving behind a Planck mass remnant
\cite{Aharonov87}. The main problem with this approach is that only
very
little energy is available to store a huge amount of information.
For a full  discussion of the criticism of this possibility,
we refer to the literature \cite{Page93}.
(3) The evolution from pure states into mixed states is an
inherent property of quantum gravity \cite{Hawking76}. This
possibility has been criticized because investigations at
the semiclassical level, which use an evolution law that is more
general
than the unitary Hamiltonian time evolution
\cite{Ellis84,Srednicki92,Liu93}, lead to problems
with energy and momentum conservation or locality
\cite{Banks84}. This is thought to be not acceptable for
a fundamental theory.

We must emphasize that in the formulation of the paradox
outlined above, only the semiclassical theory is considered. The full
problem of black hole evaporation has, however, to be treated within
quantum
gravity. The assumptions about the result of the
evaporation process rest on speculations about the outcome
of such a computation. Since a theory of quantum gravity is not yet
available, nobody has been able to
carry out such an investigation. (The attempts in two-dimensional
dilaton
gravity, reviewed
in \cite{Giddings93}, are still inconclusive.)

Recent work \cite{Ford93} has indicated that the semiclassical
Einstein
equations may well break down long before the Planck scale is
reached.
To address the above issues properly, one should thus start at least
from
a specific approach to quantum gravity, such as canonical
quantization
of general relativity.

In this Letter, we will take into account the first-order
quantum gravitational corrections to the semiclassical theory,
which are obtained by a semiclassical expansion of the Wheeler-DeWitt
equation \cite{Kiefer91}. This leads to quantum
gravitational corrections to the functional Schr\"odinger equation,
which yield
an effective non-unitarity for the
evolution of matter fields in a curved background.

How are these correction terms obtained? The starting point is the
Wheeler-DeWitt
equation
\be {\cal H}\Psi\equiv \left(-\frac{16\pi \hbar^2 G}{c^2}
 G_{abcd}\frac{\delta^2}{\delta h_{ab}\delta h_{cd}} -
\frac{c^4}{16\pi  G}\sqrt{h}R +{\cal  H}_m\right)\Psi=0, \lb{1} \ee
where $R$ is the Ricci-scalar on a three-dimensional space, and
${\cal H}_m$
denotes the Hamiltonian density for non-gravitational fields. One
then writes
the full wave functional as $\Psi\equiv \exp(iS/\hbar)$ and expands
$S$ in
powers of the gravitational constant: $S=G^{-1}S_0 + S_1 + GS_2
+\ldots$.
This ansatz is inserted into (1), which leads to equations at
consecutive
orders of $G$. The highest order yields the gravitational
Hamilton-Jacobi
equation for $S_0$, which is equivalent to Einstein's field equations
\cite{Gerlach69}. Any solution for $S_0$ gives a family of classical
spacetimes. The next order leads to the functional Schr\"odinger
equation for
non-gravitational fields on one of the background spacetimes defined
by
$S_0$ \cite{Banks85},
\be i\hbar\int d^3x G_{abcd}\frac{\delta S_0}{\delta h_{ab}}\frac{
    \delta\psi}{\delta h_{cd}}\equiv i\hbar
\frac{\partial\psi}{\partial t}
 =\int d^3x {\cal H}_m \psi. \lb{2} \ee
Here, $\psi$ denotes a wave functional which is constructed from
$S_0$ and
$S_1$ and thus
depends on the three-metric as well as non-gravitational fields. Eq.
(2)
corresponds to the level of quantum field theory on a curved
spacetime.
The Hawking radiation of an evaporating black hole is described on
this
level and has been discussed in \cite{Freese85} with the help of the
functional Schr\"odinger equation.

The next order in this expansion scheme then yields correction terms
to (2)
  since it takes into account the interaction of quantum
gravitational
fluctuations with the non-gravitational fields. The corrected
Schr\"odinger
equation (the ``first post-Schr\"odinger approximation") was derived
in \cite{Kiefer91} and reads
\be i\hbar\frac{\partial\psi}{\partial t} = \int d^3x
\left[ {\cal H}_m
+\frac{4\pi G}{c^4}\frac{{\cal H}_m^2}{\sqrt{h}R}
+ i \frac{4\pi  \hbar G}{c^4} \frac{\delta}{\delta\tau}
\left(\frac{{\cal H}_m}{\sqrt{h}R}\right)\right] \psi. \lb{3} \ee
It was shown in \cite{Kiefer91} that the correction terms are
independent of the factor ordering ambiguity inherent in (1).
The first correction term in (3) is hermitean and leads to a shift in
energy eigenvalues.
The second term is {\em non}-hermitean and thus leads to a
quantum-gravitationally induced non-unitarity in the evolution of the
wave
functional $\psi$. The presence of such a term is not surprising. It
arises simply because we have written down an effective equation for
part
of the system only and neglected degrees of freedom of the quantized
gravitational field. Similar terms can be found in QED if one
attempts a
description of the matter fields without taking into account the
degrees of
freedom of the quantized electromagnetic field \cite{Kiefer92}.
The occurrence of the non-unitary term can also be understood from
the
 fact that the Wheeler-DeWitt equation (1) leads to a Klein-Gordon
type
of conservation law. Expanding this conservation law in powers of $G$
leads to the conservation of the Schr\"odinger inner product at order
 $G^0$, which is then violated at order $G$. This violation is
reflected
 by the occurrence of the non-unitary term in (3).

The corrected Schr\"odinger equation (3) is only {\em one} equation,
in spite of the infinitely many Wheeler-DeWitt equations (1) (one at
 each space point). This happens since one has already chosen a
specific
 slicing of one of the spacetimes described by $S_0$ at the level of
(2).
The correction terms in (3) express the influence of the ``quantum
 fluctuations of spacetime" as an effective contribution to
the Schr\"odinger equation on a given spacetime.

Although the conclusions of Ref.\ \cite{Banks84} about the disturbing
implications of  a more general than Hamiltonian time evolution do
not directly
apply to the present investigation, it would not be
astonishing to find similar phenomena. It is well known
that in a semiclassical theory one must expect precisely
such kinds of inconsistencies \cite{Eppley77}.

The correction terms in (3) are formally undefined (since they
involve delta
functions at coinciding space points) and have to be regularized
before
definite physical predictions can be extracted.  Although it is not
yet clear how to perform such a regularization consistently, we argue
that qualitative predictions for the evaporation of black holes can
be
obtained from (3). Since the background spacetime in this case is
Ricci-flat,
the Ricci scalar in (3) vanishes for the considered foliation.
The derivation in \cite{Kiefer91},
however,
was performed for the case of compact three-geometries. Since we deal
here
with an asymptotically flat spacetime, we must take into account the
boundary
terms in the (integrated)  Wheeler-DeWitt equation. In the present
case this
is just the ADM energy $Mc^2$, where $M$ is the mass of the black
hole. A brief inspection of (1) thus shows that
one must replace the expression $\sqrt{h}R$ in (3) by
$-16\pi GM/c^2$, since now the ADM energy is an additional potential
 term in (1).
This is already clear from dimensional arguments, since  $ \sqrt{h}R$
has the
dimension of a length, and the only length scale in our example is
given by
the Schwarzschild radius of the black hole. One should not be
disturbed by
the fact that this replacement brings another factor of $G$ into
play. The
approximation scheme discussed in \cite{Kiefer91} is valid as long as
the
correction terms in (3) are small compared to the dominant term, even
if the
expansion is performed with some suitable small parameter different
from $G$.

We now estimate the order of magnitude of the correction
terms for
the case of the evaporating black hole. The ratio of the first,
hermitean,
correction term to the dominant term, which is ${\cal H}_m$,
is
basically given by the ratio of the energy of the field to $Mc^2$,
which
should be small even if $M$ approaches the Planck regime. The only
relevant
contribution, which we will denote by $\Delta H_m$, comes from the
second term in (3), which contains the time derivative of the mass of
the black hole:
\be \Delta H_m = -\frac{4\pi i G\hbar}{c^4}\int d^3x {\cal H}_m
\frac{\delta}{\delta\tau}\left(\frac{c^2}{16\pi GM}\right). \lb{4}
\ee
We have neglected here the time derivative of ${\cal H}_m$, since the
change of the background geometry is the dominant contribution in the
process of black
hole evaporation. To evaluate this further, one needs the time
dependence
of the mass of the black hole. We take a simple model evolution law
which
can be obtained by phenomenological arguments
(see, e.g., \cite{Witt75}):
\be M(t) = \left(M_0^3 - \frac{a\hbar c^4}{G^2}t\right)^{1/3}. \lb{5}
\ee
$M_0$ is the initial mass of the black hole, and $a$ denotes a
numerical factor which depends on the details of the model
and whose exact value is irrelevant for the present qualitative
discussion. By using (5) we of course implicitly assume that the
black hole
does not settle to a mass which is much bigger than the Planck mass.
The non-unitary contribution (4) becomes relevant, if it is of the
same
order of magnitude as the dominant term in (3), which is the
Hamiltonian
$ {\cal H}_m$ of the non-gravitational field, i.e. if
\be \frac {\hbar}{4M^2c^2}\frac{\partial M}{\partial t} \approx
\left(\frac{m_{Pl}}{M}\right)^4 \approx 1. \lb{6} \ee
The non-unitary terms thus become important if the mass of the
evaporating
black hole approaches the Planck mass. This happens after a time
$ t_* \approx \left(M_0/m_{Pl}\right)^3 t_{Pl}$.
After the mass of the black hole has entered the Planck regime, the
semiclassical expansion breaks down and the full (as yet unknown)
quantum
theory of gravity comes into play.
 Depending on the detailed
scenario,
however, the correction term (4) may be a very good approximation.
This
may happen if the black hole settles to a remnant which is of the
order
of the Planck mass but such that the numerical value of (4) is still
small
compared to ${\cal H}_m$. It may also happen that the final
evaporation
time is of the order of $M_0^4$ \cite{Carlitz87} which would mean
that (4)
is an excellent approximation for a long period of time.

What is the relevance of this violation of unitarity for the
information loss
paradox discussed above?
 For this purpose let us investigate the implications of a
non-hermitean
Hamiltonian $H_m$ for the time evolution of a
system. The density
matrix $ \rho = \sum_n p_n | \phi_n \rangle \langle
\phi_n | $ obeys the evolution equation
\be
\dot \rho = - i ( H_m \rho - \rho H_m^{\dagger}).
\label{8} \ee
The quantity $\hbox{Tr} \rho^2 - (\hbox{Tr} \rho)^2$ is
a measure for the deviation of $\rho$ from a pure state.
In the presence of a nonunitary evolution, it will in general
not remain constant in time:
\be \frac{d}{dt}
\left( \hbox{Tr}\rho^2 - (\hbox{Tr}\rho )^2
\right) = 4 \hbox{Tr} \left( (\rho^2 - \rho \hbox{Tr}
\rho) \hbox{Im} H_m \right).
\label{9} \ee
This feature is not present in ordinary
quantum mechanics: Eq.  (8) shows that for
$H_m = H_m^{\dagger}$ the deviation from a pure
state remains always constant in time. One also recognizes from
(8) that  this
mechanism does not operate if the initial state is precisely
pure with $\hbox{Tr} \rho =1$ and $\rho^2 = \rho$.
Accordingly, a pure initial state remains pure as long as our
approximation scheme is valid. The non-unitarity of (7) arises
since, although all wave functions $\phi_n$ in the decomposition of
the
density matrix obey a deterministic equation (the ``corrected
Schr\"odinger
equation" (3)), the {\em relative} norms of these wave functions can
change.
This explains why the right-hand side of (8) is zero for pure states.

It is now important to note that in the present case $\hbox{Im} H_m <
0$.
This can be recognized from (4) by taking into account that
the mass of the black hole {\em decreases}, i.e. that $\dot{M}<0$.
As a consequence, a mixed matter state will become more pure if
the correction term comes into play. One might think of this
as a partial ``recovery of information" from the black hole and,
therefore, as a hint that full quantum gravity may {\em preserve}
unitarity.

To conclude, we have discussed the first-order quantum gravitational
corrections to the functional Schr\"odinger equation for
matter fields in a curved background. Because the quantum
fluctuations of the gravitational field were neglected in
the approximation scheme, an effective non-hermitean
Hamiltonian  was found.
For the case of black hole evaporation it was found that the
non-unitary
correction terms become comparable to the unperturbed
Hamiltonian in the final stages of evaporation when
the black hole mass reaches the Planck mass. Their tendency is to
reduce the degree of mixture for a non-gravitational state.

Finally, we want to emphasize that the
unitarity-violating terms do not occur at a fundamental
level. They arose only because the quantum fluctuations
of the gravitational field  were neglected in the
expansion of the Wheeler-DeWitt equation. In a genuine
quantum gravitational treatment of the black hole
evaporation problem, unitarity may well be preserved. As we mentioned
above, the sign of the correction term indicates
that this might be the case.  It is,
however, still
an open problem \cite{Kuchar92}.

\begin{center}
{\bf Acknowledgements}
\end{center}
We are grateful to Domenico Giulini, Carlos Lousto, and H.-Dieter
Zeh for useful discussions and critical comments.
One of us (C. K.) acknowledges support from the Thomalla foundation.



\begin{thebibliography}{99}
\bibitem{Hawking76} S. W. Hawking, Phys. Rev. D {\bf 14}, 2460
(1976).

\bibitem{Page93} D. N. Page, Preprint hep-th 9305040
(1993), to be published in the {\em Proceedings of the 5th
Canadian Conference on General Relativity and Relativistic
Astrophysics 1993}.

\bibitem{Preskill92} J. Preskill, Caltech Report CALT-68-1819,
hep-th 9209058 (1992).

\bibitem{Page80} D. N. Page, Phys. Rev. Lett.
{\bf 44}, 301 (1980).

\bibitem{Mueller93} R. M\"uller and C. O. Lousto, Universit\"at
Konstanz Report, gr-qc 9307001 (1993).

\bibitem{Bekenstein93} J. D. Bekenstein, Phys. Rev. Lett.
{\bf 70}, 3680 (1993).

\bibitem{Schiffer93} M. Schiffer, CERN Report CERN-TH. 6811/93,
hep-th 9303011 (1993).

\bibitem{Carlitz87} R. D. Carlitz and R. S. Willey, Phys. Rev.
D {\bf 36}, 2336 (1987).

\bibitem{Wilczek93} F. Wilczek, Princeton Report IASSNS-HEP-93/12,
 hep-th/9302096 (1993).

\bibitem{Aharonov87} Y. Aharonov, A. Casher, and
S. Nussinov, Phys. Lett. B {\bf 191}, 51 (1987).

\bibitem{Ellis84} J. Ellis, J. S. Hagelin, D. V. Nanopoulos,
and M. Srednicki, Nucl. Phys. B {\bf 241}, 381 (1984).

\bibitem{Srednicki92} M. Srednicki, University of California, Santa
Barbara
Report UCSB-TH-92-22, hep-th 9206056 (1992).

\bibitem{Liu93} J. Liu, Stanford University Report SU-ITP-93-1,
hep-th 9301082 (1993).

\bibitem{Banks84} T. Banks, L. Susskind, and M. E. Peskin,
Nucl. Phys. B {\bf 244}, 125 (1984).

\bibitem{Giddings93} S. Giddings, University of California, Santa
Barbara
Report UCSB-TH-93-16, hep-th 9306041 (1993).

\bibitem{Ford93} C.-I. Kuo and L. H. Ford, Phys. Rev. D {\bf 47},
4510 (1993); T. Padmanabhan and T. P. Singh, Class. Quantum Grav.
{\bf 7}, 411 (1990).

\bibitem{Kiefer91} C. Kiefer and T. P. Singh, Phys. Rev. D {\bf 44},
1067 (1991).

\bibitem{Gerlach69} U. H. Gerlach, Phys. Rev. {\bf 177}, 1929 (1969).

\bibitem{Banks85} T. Banks, Nucl. Phys. B {\bf 249}, 332 (1985).

\bibitem{Freese85} K. Freese, C. T. Hill, and M. Mueller, Nucl. Phys.
B {\bf 255}, 693 (1985).

\bibitem{Kiefer92} C. Kiefer, Phys. Rev. D {\bf 45}, 2044 (1992).

\bibitem{Eppley77} K. Eppley and E. Hannah, Found. Phys.
{\bf 7}, 51 (1977).

\bibitem{Witt75} B. S. DeWitt, Phys. Rep. {\bf 19}, 295 (1975),
R. M. Wald, {\em General Relativity} (University of Chicago Press,
1984).

\bibitem{Kuchar92}  K. V. Kucha\v{r}, in {\em Proceedings of the 4th
Canadian
Conference on General Relativity and Relativistic Astrophysics}, ed.
by
G. Kunstatter, D. Vincent and J. Williams (World Scientific,
Singapore, 1992).


\end{thebibliography}
\end{document}